\documentstyle[sprocl,epsfig,psfig]{article}
\begin{document}

\vspace*{-25mm}
\hfill MZ-TH/99-34\\[-3ex]

\hfill WUE-ITP-99-016

\vspace{12mm}

\title{LEPTOQUARK AND $R$-PARITY VIOLATING SUSY PROCESSES \footnote{Talk
    given by R.\ R\"uckl in the working group session P6 at the
    International Workshop on Linear Colliders, Sitges, Barcelona,
    Spain, April 28 - May 5, 1999.}}

\author{M.\ HEYSSLER, R.\ R\"UCKL} 

\address{Institut f\"ur Theoretische Physik, Universit\"at 
W\"urzburg, D--97074 W\"urzburg}

\author{H.\ SPIESBERGER}

\address{Institut f\"ur Physik, Johannes-Gutenberg-Universit\"at
  Mainz, D--55099 Mainz}

\maketitle\abstracts{Pair production of leptoquarks at future $e^+e^-$
  linear colliders is investigated for center-of-mass energies of
  500~GeV and 800~GeV and an integrated luminosity of $500$~fb$^{-1}$.
  Based on a Monte Carlo simulation we estimate the event rates for
  signal as well as background processes and evaluate the discovery
  potential.  As an example of virtual effects we consider deviations
  from standard model predictions due to $R$-parity violating sneutrino
  exchange in purely leptonic processes.}


\section{Introduction}

If leptoquarks (LQ) exist with sufficiently low masses, they would
affect measurements at present and future colliders in many ways. In
electron-positron collisions they may be produced in pairs or as single
particles, while virtual LQ exchange may show up in $e^+e^- \rightarrow$
hadrons. In this study, we update a previous investigation \cite{RSS96}
of LQ pair production focussing on the possibility of an integrated
luminosity as high as $500$~fb$^{-1}$, while in Ref.\ \cite{RSS96} much
lower luminosities, $20$~fb$^{-1}$ at the center-of-mass energy
$\sqrt{s}=500$ GeV and $50$~fb$^{-1}$ at $\sqrt{s}=800$ GeV are
considered.

Specifically, in SUSY models with $R$-parity violating Yukawa couplings
one can have processes of the above kind with squarks playing the role
of leptoquarks. Moreover, pair or single production of sleptons and
slepton exchange can give rise to purely leptonic processes.  Here, we
illustrate the impact of virtual effects for the case of sneutrino
exchange in $e^+e^- \rightarrow l^+l^-$ where $l=e,\mu,\tau$
investigated previously for LEP2 \cite{KRSZ97}.

References to related work in the literature can be found in 
Ref.\ \cite{RSS96,KRSZ97}.

\section{LQ Pair Production}

\begin{figure}[htbp] 
\unitlength 1mm
\begin{picture}(120,20)
\put(30,-3){
\epsfig{file=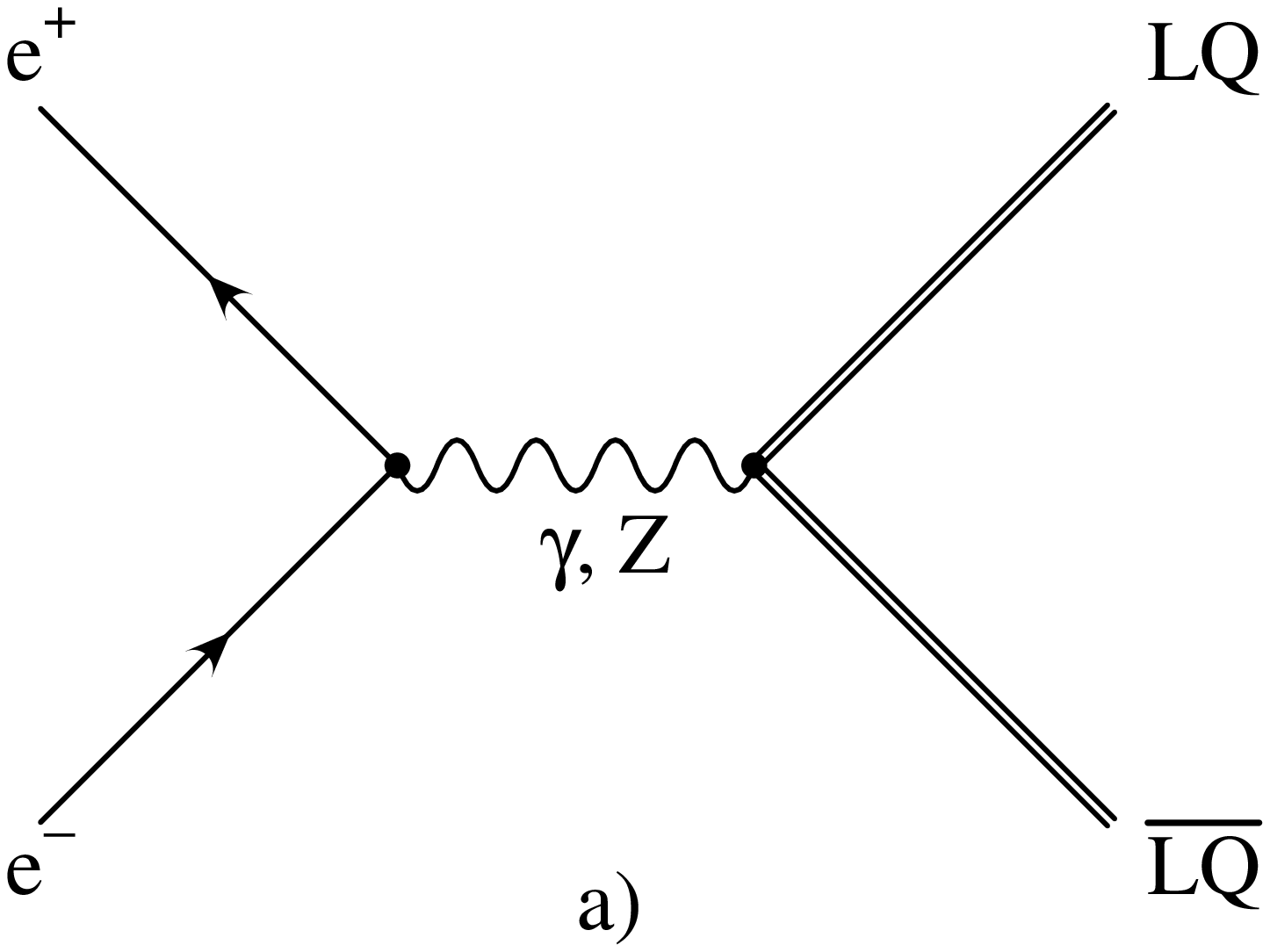,height=2.4cm,width=3.4cm}}
\put(75,-4){
\epsfig{file=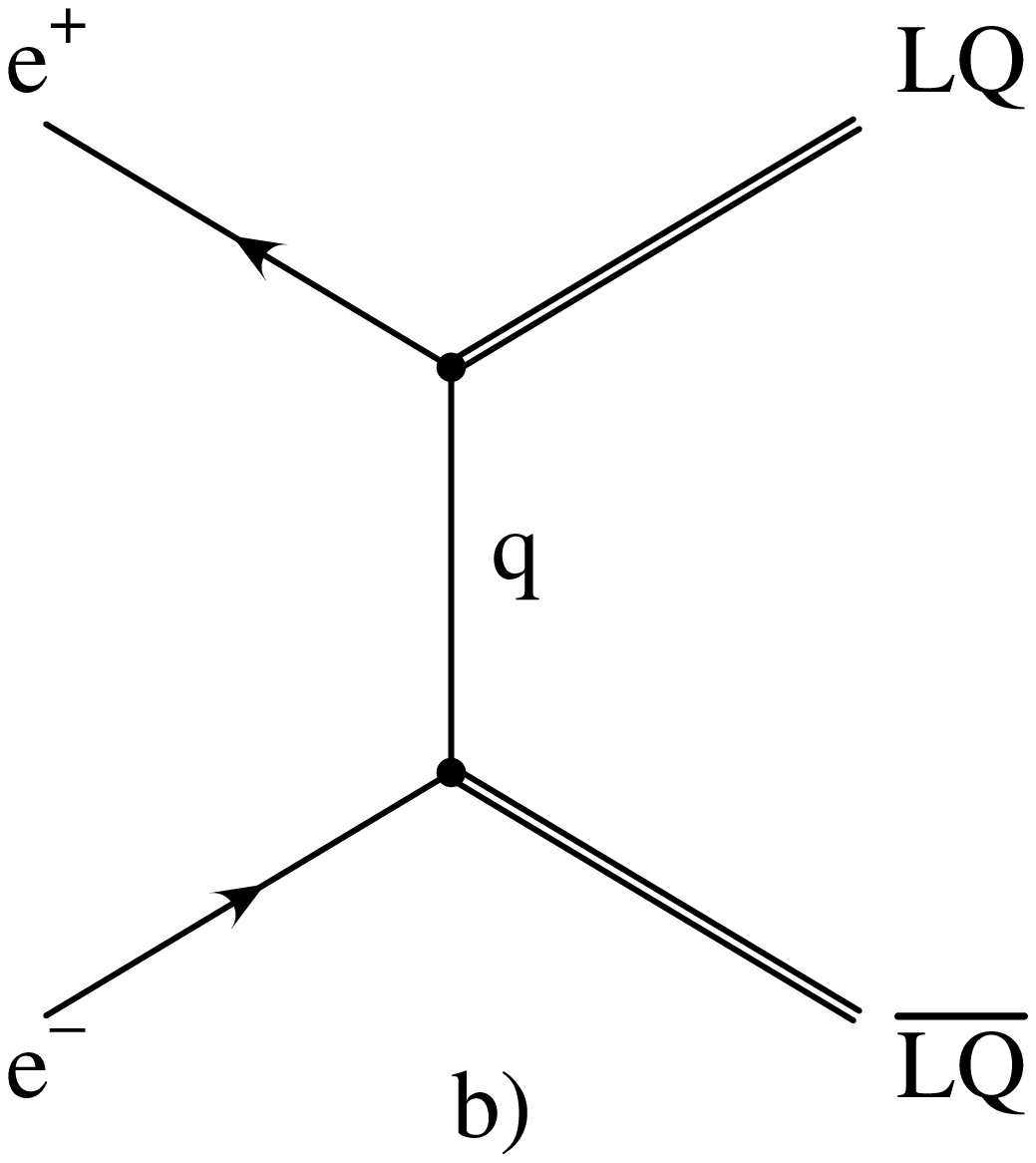,height=2.6cm,width=2.3cm}}
\end{picture}
\caption{\it \footnotesize LQ pair production in $e^+e^-$ colliders at
  lowest order.} 
\label{fig:Fgraphs}
\end{figure}

The general theoretical framework of LQ quantum numbers and couplings is
described in Ref.\ \cite{BRW87,BR93}. Following the usual procedure we
assume that only a single multiplet of mass degenerate leptoquarks is
present at a time and that these decay only into standard model leptons
and quarks.  Furthermore, we restrict ourselves to leptoquarks of the
first generation decaying into $e^{\pm}$ or $\nu_e$ plus a jet.  In
lowest order, LQ pairs are produced in $e^+e^-$ annihilation via
$\gamma$ and $Z$ exchange (Fig.\ \ref{fig:Fgraphs}a) and via quark
exchange (Fig.\ \ref{fig:Fgraphs}b).  In the first case, the amplitude
is determined by LQ gauge couplings, i.e.\ their electric charge and
weak isospin, whereas the second case involves LQ-lepton-quark Yukawa
couplings $\lambda$. Experimental constraints \cite{bounds} require the
latter to be approximately chiral.  At $\sqrt{s} = 500$ GeV and for
negligible $\lambda$, the production cross sections \cite{RSS96} range
from 6 -- 150 fb for scalar leptoquarks of mass $m_{\rm LQ} = 200$ GeV
and from 0.2 -- 1.8 pb for vector leptoquarks. Beamsstrahlung and
initial state radiation are included in these estimates.

\subsection{Monte Carlo simulation}

We have performed a detailed Monte Carlo study based on the event
generator {\tt LQPAIR} \cite{hubert96} with the lowest order matrix
elements are taken from Ref.~\cite{BR93} and corrected for beamstrahlung
and initial state radiation.  Hadronic final states are generated by an
interface to the {\sc Lund} Monte Carlo programs for jet fragmentation.
In the event analysis we have used the Durham jet algorithm. Finally, an
interface to the program {\tt SMEAR} \cite{smear96} allows for a
realistic detector simulation.

The LQ decays give rise to three different event topologies called I to
III in Table \ref{tab:cuts} without/with missing momentum
$p^{\mathrm{miss}}$ carried away by neutrinos.  For clean event
identification and precise mass reconstruction, cuts on energies and
transverse momenta have to be applied.  The most relevant of them are
given in Table \ref{tab:cuts}. For further details we refer to
Ref.~\cite{RSS96}.


\begin{table}[ht]
\begin{center}
\begin{tabular}{|c|c|c|c|}
\hline
variable & (I) $e^+e^- + 2$ jets & (II) $e^\pm + 2$ jets $ +
  p^{\mathrm{miss}} $ & (III) 2 jets  $ + p^{\mathrm{miss}} $ \\\hline
$ p^e_T $ & $ \ge 20 $ GeV & $ \ge 20 $ GeV & $ \ge 20 $ GeV \quad veto\\
$ p^{\mathrm{miss}}_T $ & $ \le 25 $ GeV & $ \ge $ 25 GeV & $ \ge $ 25 GeV \\
$ E_j $ & $ \ge 10 $ GeV & $ \ge 10 $ GeV & --\\ 
$ E_{\mathrm{vis}} $ & $ \ge 0.9 \sqrt{s} $ & $ \ge 0.6 \sqrt{s} $ & --\\ 
$ p^j_T $ & -- & -- & $ \ge 75 $ GeV \\
$ E_{\mathrm{had}} $ & -- & $ \ge 150 $ GeV & $ \le 300 $ GeV\\
\hline
\end{tabular}
\end{center}
\caption[]{\it \footnotesize General cuts used for event
  identification.} 
\label{tab:cuts}
\end{table}


\subsection{Background Processes}

The dominant standard model background is due to vector boson pair
production, $e^+e^- \rightarrow Z Z,~W^+ W^-$, with one $Z$ or $W$
decaying leptonically and the other one decaying into two jets. This
background is simulated with the help of {\tt WPHACT} \cite{wphact}
including single- and non-resonant four-fermion production. Another
significant source for signal contamination is top quark production,
$e^+e^- \rightarrow t \bar t$, followed by $t \rightarrow b W$, 
with the $W$ decaying leptonically.  In order to suppress the above
background the following additional cuts are applied \cite{RSS96}:
\begin{itemize}
\item[] 
$|M_{\ell_1 \ell_2} - M_{Z,W}| \geq 10$~GeV (I), ~~~ 20 GeV (II,III), 
\item[]
$|M_{j_1 j_2} - M_{Z,W}|       \geq 10$~GeV (I), ~~~ 20 GeV (II,III), 
\item[]
$M_{\ell_1 \ell_2}$, 
$M_{\ell_1 j_2}$, 
$M_{j_1 \ell_2}                \geq 20$~GeV (I,II), 
~~~$M_{j_1 j_2}   \leq 400$~GeV (III).  
\end{itemize}
The indices $j_1$ and $j_2$ label the two jets ordered by their
energies such that $E_{j_1} > E_{j_2}$, whereas $\ell_1$ and $\ell_2$ 
label the leptons ordered by
$|M_{\ell_1 j_1} - M_{\ell_2 j_2}| < |M_{\ell_1 j_2} - M_{\ell_2 j_1}|$.

The number of background events in channel (I) to (III) surviving the
above cuts at $\sqrt{s}=500~(800)$ GeV and $500$~fb$^{-1}$ integrated
luminosity is obtained from Table~5 of Ref.~\cite{RSS96} after
multiplication by factors of 25~(10) for $\sqrt{s}=500~(800)$ GeV to
account for the higher luminosity.  Similarly, one can infer the number
of LQ events expected for $500$~fb$^{-1}$ after cuts from the
corresponding numbers for $20$~fb$^{-1}$ and $50$~fb$^{-1}$,
respectively, given in Table 6 of Ref.~\cite{RSS96}.  There the rates  
in channel (I) and (II) are compared for the vector ($V_0$) and scalar
($S_0$) states with the least favourable production cross sections, and
for two collider energies choosing $m_{LQ} \approx 0.9\sqrt{s}$.  For
channel (I), results are also given for $V_1$ and $S_1$ having the
largest production cross sections among the vector and scalar states,
respectively.  The apparent strong dependence of the event rates on the
LQ quantum numbers together with a mass measurement should allow for an
efficient discrimination of different LQ hypotheses.


\subsection{Sensitivity limits}

A rough  estimate of the sensitivity limits can be based on the total
number of events.  To that end we determine the range of leptoquark
masses, for which the number of signal events is equal to or larger than
five times the number $N_{bg}$ of background events.  The upper limits
of these mass ranges are summarized in Table~\ref{tab:massreach}
together with the expection for the lower luminosities assumed in
Ref.~\cite{RSS96}. Note that the background in channel I is
approximately 10~(200) times smaller than the background in channel
II~(III).

Accepting the requirement of a $5\sigma$ effect as a sensible discovery
criterion, one finds that at $\sqrt{s}=500~(800)$ GeV and with
$500$~fb$^{-1}$, scalar leptoquarks can be discovered for masses up to
86--98\% ~(50--99\%) of $\sqrt{s}/2$, while vector states should be
observable for masses up to 97--99\% (86--99\%) of $\sqrt{s}/2$.  The
corresponding mass limits for $\sqrt{s}=500$ GeV and $20$~fb$^{-1}$~
($\sqrt{s}=800$ GeV and $50$~fb$^{-1}$) are 73--98\% (37--97\%) of
$\sqrt{s}/2$ for scalars, and 94--99\% (82--99\%) of $\sqrt{s}/2$ for
vectors.  The higher luminosity thus helps to increase the mass reach
in the case of scalar leptoquarks with less favourable quantum numbers
by about (0.1 to  0.2) $\sqrt{s}/2$. In particular, the species $^{-1/3}S_0$
and $^{-1/3}S_1$ which are unobservable in channel II for the lower
luminosities and for masses above 100 GeV can be probed with the high
luminosity in an interesting mass range.  The same holds for
$^{-2/3}S_{1/2}$ in channel III at $\sqrt{s}=800$ GeV.


\section{Sneutrino Exchange}

\begin{figure}[htbp] 
\unitlength 1mm
\begin{picture}(125,20)
\put(3,-3){
\epsfig{file=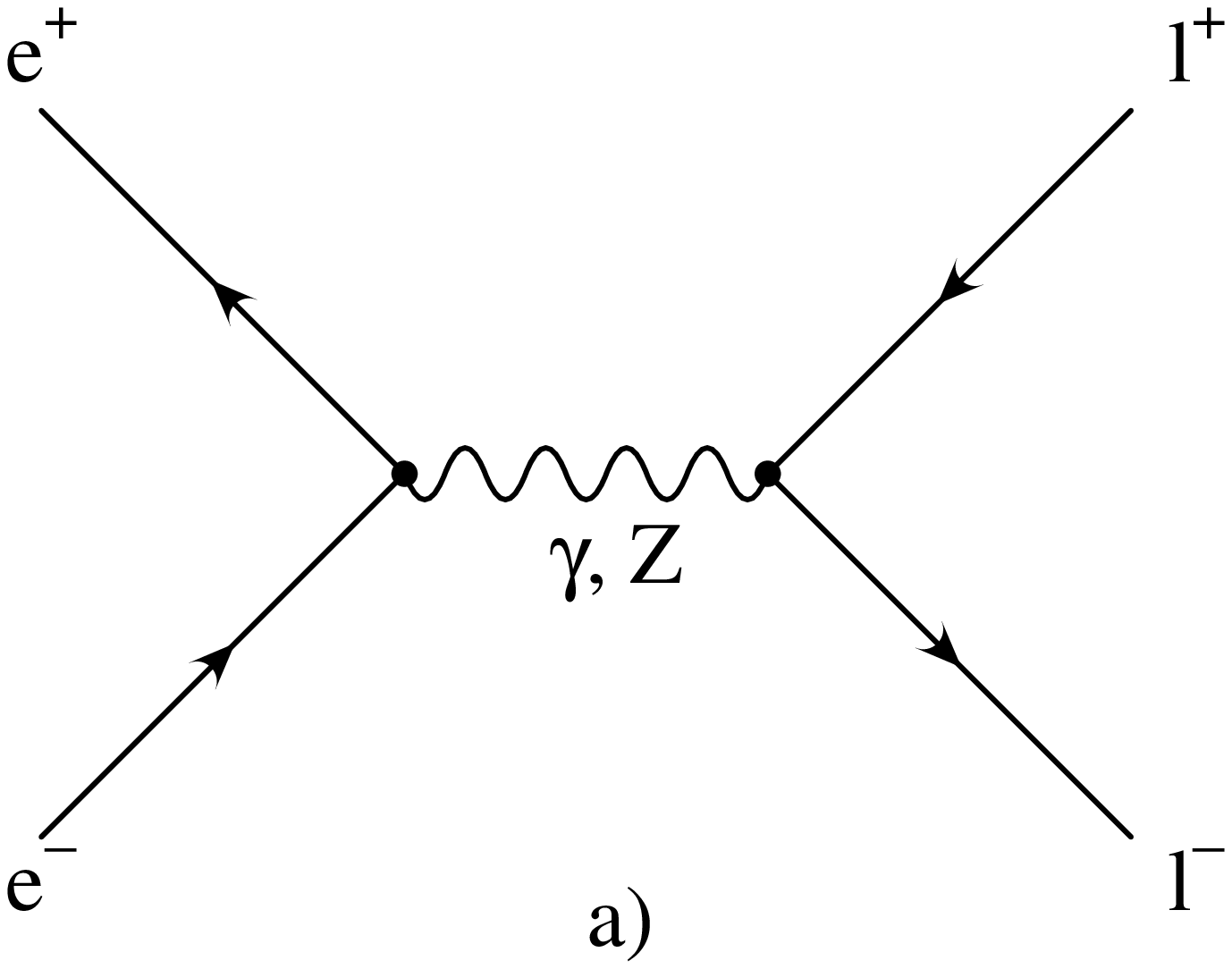,height=2.4cm,width=3.4cm}}
\put(41,-4){
\epsfig{file=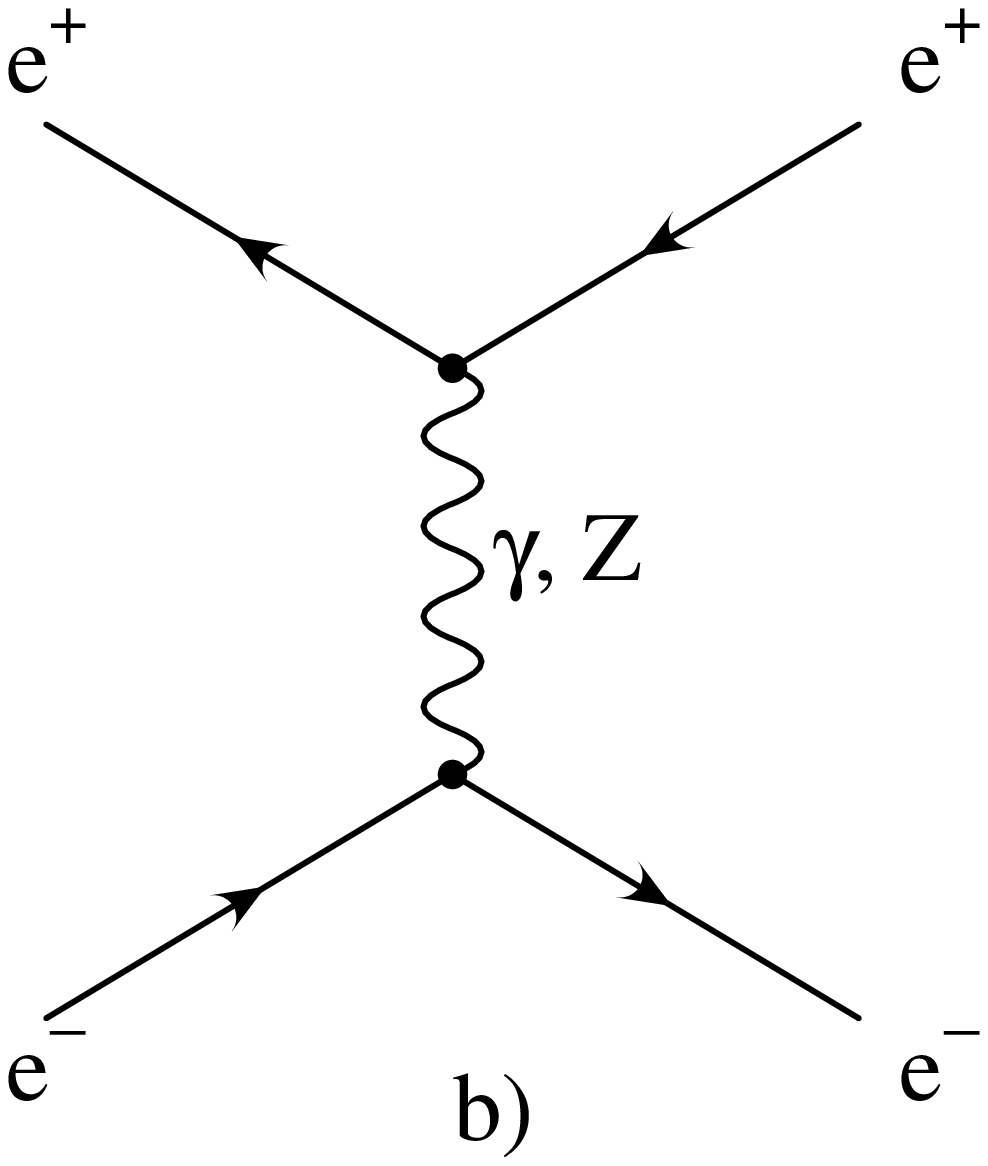,height=2.6cm,width=2.3cm}}
\put(67,-3){
\epsfig{file=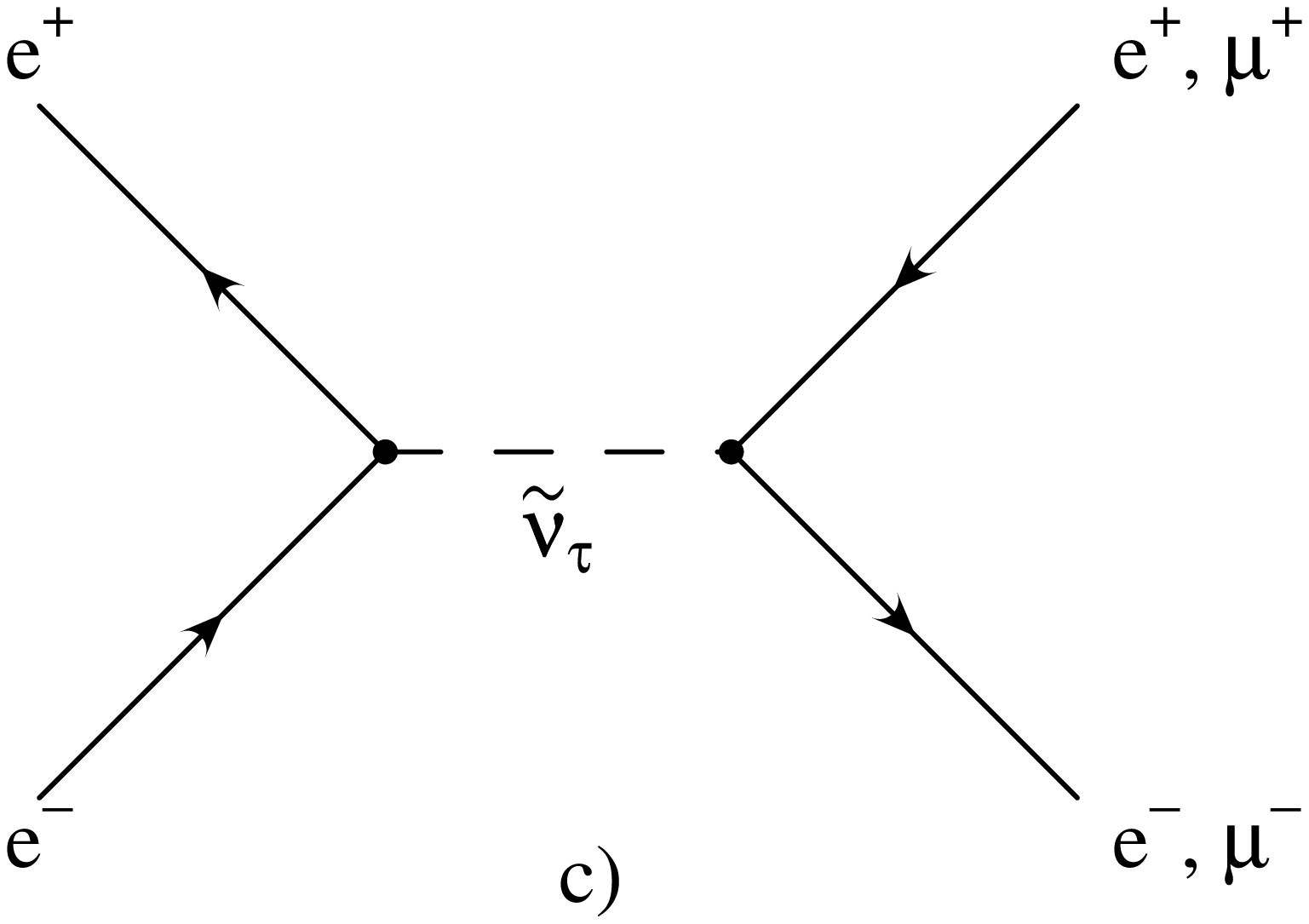,height=2.4cm,width=3.4cm}}
\put(103,-4){
\epsfig{file=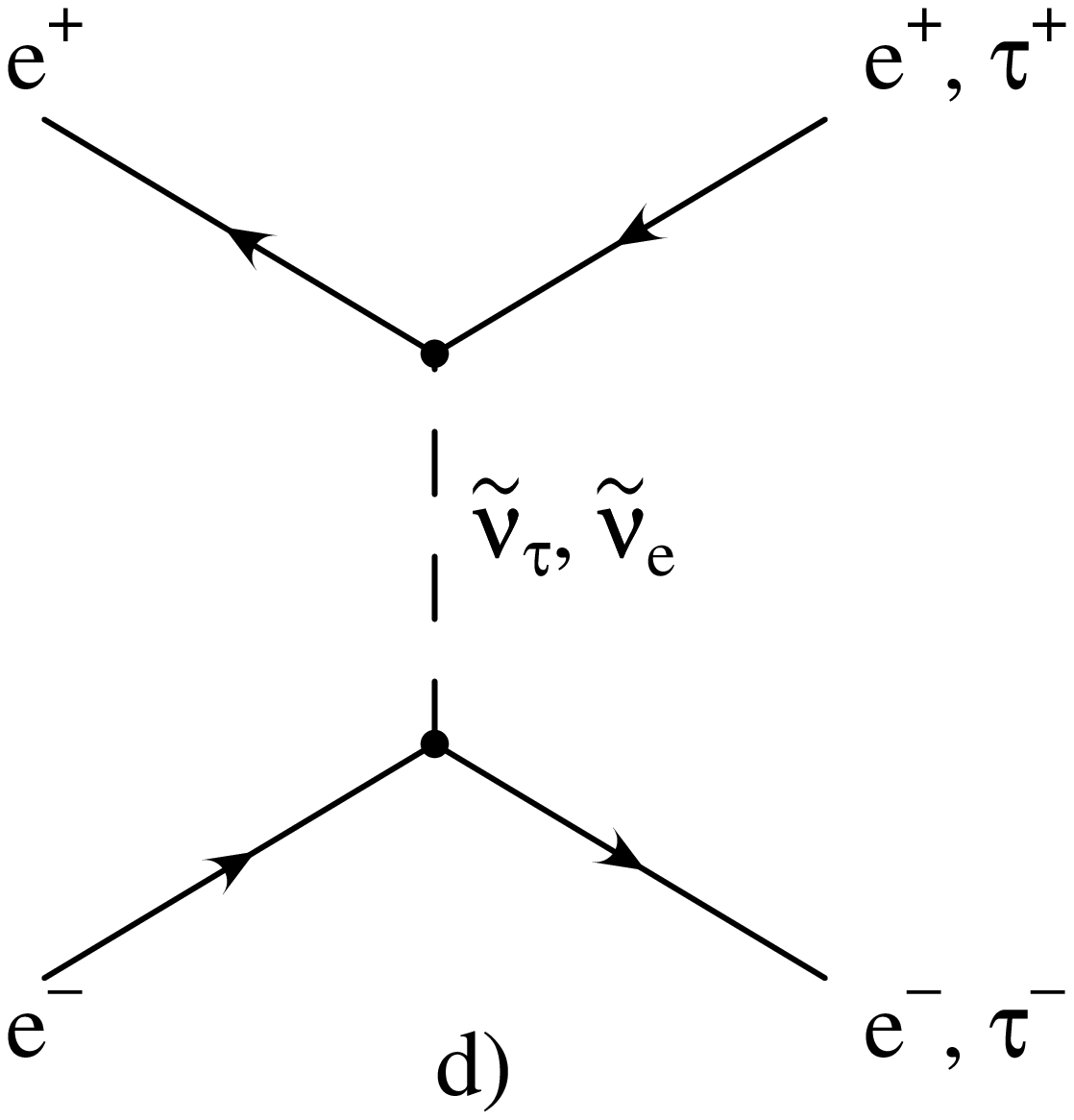,height=2.6cm,width=2.3cm}}
\end{picture}
\caption{\it \footnotesize Lowest order contributions to
  $e^+e^- \rightarrow l^+l^-$ including sneutrino exchange.} 
\label{fig:Fexchange}
\end{figure}

The $R$-parity violating term $\frac{1}{2} \lambda_{ijk} L_i L_j E^c_k$
in the superpotential, where $L_i$ and $E_k$ denote left-handed doublets
and right-handed singlets of lepton superfields and $i,j,k$ are
generation indices, induces new contributions to $e^+e^- \rightarrow
l^+l^-$ from virtual sneutrino exchange as shown in Fig.\ 
\ref{fig:Fexchange}.

\clearpage

\begin{table}[t]
{\footnotesize
\setlength{\tabcolsep}{0.5mm}
\begin{center}
\rotatebox[origin=cc]{90}{
\begin{tabular}{|r|c|cc|cc|cc|}\hline
\multicolumn{8}{|c|}{\rule[-2mm]{0mm}{6mm}
  \bf B:~~ \boldmath$\sqrt{s} = 800$\unboldmath GeV,~~ 
  (50 fb\boldmath$^{-1}$\unboldmath) 
  500 fb\boldmath$^{-1}$\unboldmath} 
\\ \hline
\multicolumn{2}{|r|}{\bf search} &
\multicolumn{2}{|c|}{ \bf I}   & 
\multicolumn{2}{|c|}{ \bf II}  & 
\multicolumn{2}{|c|}{ \bf III} \\ \hline 
\multicolumn{2}{|r|}{\rule[-1mm]{0mm}{5mm}
 \boldmath$5\,\sqrt{N_{bg}}$\unboldmath} &
{ (21) }  & { 66  } & 
{ (60) }  & { 189 } & 
{ (375)}  & { 1186} \\ \hline
 \bf states & \boldmath$B_{eq}$\unboldmath &
\multicolumn{6}{|c|}{\bf  mass reach in GeV} \\ \hline\hline
{  \boldmath$^{-1/3}S_0$\unboldmath} & $^2/_3$       &
{  (318) }  & { 332  } &
{  ($\Diamond$) }    & { {\bf 311}  } &
{  ($\Diamond$) }    & { $\Diamond$  }   \\ 
 & $^1/_2$ &
{  (289) }  & { {\bf 323}  } &
{  ($\Diamond$) }    & { {\bf 309}  } &
{  ($\Diamond$) }    & { $\Diamond$  }   \\ 
 & $1$ &
{  (350) }  & { 362  } &
{  (-) }    & { -  }   &
{  (-) }    & { -  }   \\ \hline
{\rule[-1mm]{0mm}{4.5mm}
  \boldmath$^{-4/3}\widetilde{S}_0$\unboldmath} & $1$ &
{  (387) }  & { 391  } &
{  (-) }    & { -  } &
{  (-) }    & { -  }   \\ \hline
{  \boldmath$^{2/3}S_1$\unboldmath} & $0$ &
{  (-) }    & { -  } &
{  (-) }    & { -  } &
{  (275) }  & { 302  }   \\ \hline
{  \boldmath$^{-1/3}S_1$\unboldmath} & $^1/_2$ &
{  (289) }    & { {\bf 323}  } &
{  ($\Diamond$) }    & { {\bf 311}  } &
{  ($\Diamond$) }    & { $\Diamond$  }   \\ \hline
{  \boldmath$^{-4/3}S_1$\unboldmath} & $1$ &
{  (389) }    & { 396  } &
{  (-) }    & { -  } &
{  (-) }    & { -  }   \\ \hline
{  \boldmath$^{-2/3}S_{1/2}$\unboldmath} & $^1/_2$ &
{  (369) }    & { 385  } &
{  (359) }    & { 377  } &
{  ($\Diamond$) }    & { {\bf 308} }   \\ 
 & $0$ &
{  (-) }    & { -  } &
{  (-) }    & { -  } &
{  (239) }    & { {\bf 287} }   \\ 
 & $1$ &
{  (384) }    & { 394  } &
{  (-) }    & { -  } &
{  (-) }    & { -  }   \\ \hline
{  \boldmath$^{-5/3}S_{1/2}$\unboldmath} & $1$ &
{  (389) }    & { 396  } &
{  (-) }    & { - } &
{  (-) }    & { - }   \\ \hline
{\rule[-1mm]{0mm}{4.5mm}
  \boldmath$^{1/3}\widetilde{S}_{1/2}$\unboldmath} & $0$ &
{  (-) }    & { -  } &
{  (-) }    & { - } &
{  (146) }  & { {\bf 198} }   \\ \hline
{\rule[-1mm]{0mm}{4.5mm}
 \boldmath$^{-2/3}\widetilde{S}_{1/2}$\unboldmath} & $1$ &
{  (379) }    & { 396  } &
{  (-) }    & { - } &
{  (-) }  & { - }   \\ \hline\hline
{  \boldmath$^{-1/3}V_{1/2}$\unboldmath} & $^1/_2$ &
{  (385) }    & { 396 } &
{  (380) }    & { 392 } &
{  (266) }    & { {\ bf 302} } \\ 
 & $0$ &
{  (-) }    & { - } &
{  (-) }    & { - } &
{  (326) }    & { 345 } \\ 
 & $1$ &
{  (392) }    & { 396 } &
{  (-) }    & { - } &
{  (-) }    & { - } \\ \hline
{  \boldmath$^{-4/3}V_{1/2}$\unboldmath} & $1$ &
{  (395) }  & { 395 } &
{  (-) }    & { - } &
{  (-) }    & { - } \\ \hline
{\rule[-1mm]{0mm}{4.5mm}
 \boldmath$^{2/3}\widetilde{V}_{1/2}$\unboldmath} & $0$ &
{  (-) }    & { - } &
{  (-) }    & { - } &
{  (326) }  & { 345 } \\ \hline
{\rule[-1mm]{0mm}{4.5mm}
 \boldmath$^{-1/3}\widetilde{V}_{1/2}$\unboldmath} & $1$ &
{  (390) }  & { 392 } &
{  (-) }    & { - } &
{  (-) }    & { - } \\ \hline
{  \boldmath$^{-2/3}{V}_{0}$\unboldmath} & $^2/_3$ &
{  (385) }    & { 392 } &
{  (373) }    & { 389 } &
{  (200) }    & { {\bf 279} } \\ 
 & $^1/_2$ &
{  (380) }    & { 392 } &
{  (376) }    & { 390 } &
{  (244) }    & { {\bf 317} } \\ 
 & $1$ &
{  (390) }  & { 392 } &
{  (-) }    & { - } &
{  (-) }    & { - } \\ \hline
{\rule[-1mm]{0mm}{4.5mm}
 \boldmath$^{-5/3}\widetilde{V}_{0}$\unboldmath} & $1$ &
{  (396) }  & { 397 } &
{  (-) }    & { - } &
{  (-) }    & { - } \\ \hline 
{  \boldmath$^{1/3}{V}_{1}$\unboldmath} & $0$ &
{  (-)  } & { -   } &
{  (-)  } & { -   } &
{  (352)} & { 376 } \\ \hline 
{  \boldmath$^{-2/3}{V}_{1}$\unboldmath} & $^1/_2$ &
{  (380) } & { 392 } &
{  (375) } & { 390 } &
{  (244) } & { {\bf 317} } \\ \hline 
{  \boldmath$^{-5/3}{V}_{1}$\unboldmath} & $^1/_2$ &
{  (396) } & { 395 } &
{  (-)   } & { - } &
{  (-)   } & { - } \\ \hline 
\end{tabular}
}    
\end{center}
}    
\end{table}


\begin{table}[b]
{\footnotesize
\setlength{\tabcolsep}{0.5mm}
\begin{center}
\rotatebox[origin=cc]{90}{
\begin{tabular}{|r|c|cc|cc|cc|}\hline
\multicolumn{8}{|c|}{\rule[-2mm]{0mm}{6mm}
  \bf A:~~ \boldmath$\sqrt{s} = 500$\unboldmath GeV,~~ 
  (20 fb\boldmath$^{-1}$\unboldmath) 
  500 fb\boldmath$^{-1}$\unboldmath} 
\\ \hline
\multicolumn{2}{|r|}{\bf search} &
\multicolumn{2}{|c|}{  \bf I}   & 
\multicolumn{2}{|c|}{  \bf II}  & 
\multicolumn{2}{|c|}{  \bf III} \\ \hline 
\multicolumn{2}{|r|}{\rule[-1mm]{0mm}{5mm}
  \boldmath$5\,\sqrt{N_{bg}}$\unboldmath} &
  { (18) }  & { 90  } & 
  { (61) }  & { 306 } & 
  { (251)}  & { 1259} \\ \hline
  \bf states & \boldmath$B_{eq}$\unboldmath &
\multicolumn{6}{|c|}{\bf   mass reach in GeV} \\ \hline\hline
{ \boldmath$^{-1/3}S_0$\unboldmath} & $^2/_3$ &
  { (202) }  & { {\bf 228}  } &
  { ($\Diamond$) }    & { {\bf 205} } &
  { ($\Diamond$) }    & { $\Diamond$   }   \\ 
 & $^1/_2$ &
  { (183) }  & { {\bf 220}  } &
  { ($\Diamond$) }    & { {\bf 208}  } &
  { ($\Diamond$) }    & { $\Diamond$   }   \\ 
 & $1$ &
{  (217) }  & { 235  } &
{  (-) }    & { -  }   &
{  (-) }    & { -  }   \\ \hline
{\rule[-1mm]{0mm}{4.5mm}
 \boldmath$^{-4/3}\widetilde{S}_0$\unboldmath} & $1$ &
{  (242) }  & { 245  } &
{  (-) }    & { -  } &
{  (-) }    & { -  }   \\ \hline
{   \boldmath$^{2/3}S_1$\unboldmath} & $0$ &
{  (-) }    & { -  } &
{  (-) }    & { -  } &
{  (225) }    & { 238  }   \\ \hline
{   \boldmath$^{-1/3}S_1$\unboldmath} & $^1/_2$ &
{  (183) }    & { {\bf 220}  } &
{  ($\Diamond$) }    & { {\bf 208}  } &
{  ($\Diamond$) }    & { $\Diamond$   }   \\ \hline
{   \boldmath$^{-4/3}S_1$\unboldmath} & $1$ &
{  (244) }    & { 245  } &
{  (-) }    & { -  } &
{  (-) }    & { -  }   \\ \hline
{   \boldmath$^{-2/3}S_{1/2}$\unboldmath} & $^1/_2$ &
{  (230) }    & { 241  } &
{  (221) }    & { 236  } &
{  (179) }    & { {\bf 212}  }   \\ 
 & $0$ &
{  (-) }    & { -  } &
{  (-) }    & { -  } &
{  (218) }    & { 235  }   \\ 
 & $1$ &
{  (240) }    & { 245  } &
{  (-) }    & { -  } &
{  (-) }    & { -  }   \\ \hline
{   \boldmath$^{-5/3}S_{1/2}$\unboldmath} & $1$ &
{  (244) }    & { 245  } &
{  (-) }    & { - } &
{  (-) }    & { - }   \\ \hline
{\rule[-1mm]{0mm}{4.5mm}
   \boldmath$^{1/3}\widetilde{S}_{1/2}$\unboldmath} & $0$ &
{  (-) }    & { -  } &
{  (-) }    & { - } &
{  (198) }  & { 214 }   \\ \hline
{\rule[-1mm]{0mm}{4.5mm}
  \boldmath$^{-2/3}\widetilde{S}_{1/2}$\unboldmath} & $1$ &
{  (237) }    & { 244  } &
{  (-) }    & { - } &
{  (-) }  & { - }   \\ \hline\hline
{   \boldmath$^{-1/3}V_{1/2}$\unboldmath} & $^1/_2$ &
{  (241) }    & { 244 } &
{  (237) }    & { 242 } &
{  (220) }    & { 237 } \\ 
 & $0$ &
{  (-) }    & { - } &
{  (-) }    & { - } &
{  (236) }    & { 242 } \\ 
 & $1$ &
{  (245) }    & { 246 } &
{  (-) }    & { - } &
{  (-) }    & { - } \\ \hline
{   \boldmath$^{-4/3}V_{1/2}$\unboldmath} & $1$ &
{  (247) }  & { 247 } &
{  (-) }    & { - } &
{  (-) }    & { - } \\ \hline
{\rule[-1mm]{0mm}{4.5mm}
 \boldmath$^{2/3}\widetilde{V}_{1/2}$\unboldmath} & $0$ &
{  (-) }    & { - } &
{  (-) }    & { - } &
{  (236) }  & { 242 } \\ \hline
{\rule[-1mm]{0mm}{4.5mm}
  \boldmath$^{-1/3}\widetilde{V}_{1/2}$\unboldmath} & $1$ &
{  (244) }  & { 245 } &
{  (-) }    & { - } &
{  (-) }    & { - } \\ \hline
{   \boldmath$^{-2/3}{V}_{0}$\unboldmath} & $^2/_3$ &
{  (241) }    & { 244 } &
{  (233) }    & { 241 } &
{  (195) }    & { {\bf 220} } \\ 
 & $^1/_2$ &
{  (238) }    & { 242 } &
{  (234) }    & { 239 } &
{  (212) }    & { 227 } \\ 
 & $1$ &
{  (244) }  & { 247 } &
{  (-) }    & { - } &
{  (-) }    & { - } \\ \hline
{\rule[-1mm]{0mm}{4.5mm}
  \boldmath$^{-5/3}\widetilde{V}_{0}$\unboldmath} & $1$ &
{  (247) }  & { 247 } &
{  (-) }    & { - } &
{  (-) }    & { - } \\ \hline 
{   \boldmath$^{1/3}{V}_{1}$\unboldmath} & $0$ &
{  (-)  } & { -   } &
{  (-)  } & { -   } &
{  (241)} & { 245 } \\ \hline 
{   \boldmath$^{-2/3}{V}_{1}$\unboldmath} & $^1/_2$ &
{  (238) } & { 242 } &
{  (234) } & { 239 } &
{  (212) } & { 227 } \\ \hline 
{   \boldmath$^{-5/3}{V}_{1}$\unboldmath} & $^1/_2$ &
{  (248) } & { 247 } &
{  (-)   } & { - } &
{  (-)   } & { - } \\ \hline 
\end{tabular}
}    
\end{center}
}    
\caption[]{\it \footnotesize Number of signal events required for a
  $5\sigma$ effect in channel I to III and corresponding mass reach in
  GeV for all leptoquark states classified in Ref.~\cite{BRW87} and for
  the cuts described in the text. Significant gain due to luminosity is
  emphasized by boldfaced numbers. $B_{eq}$ denotes the LQ branching ratio
  into a charged lepton plus jet. Search channels not accessible are
  marked by a dash, sensitivity limits below 100~GeV by a diamond.} 
\label{tab:massreach}
\end{table}

\clearpage

\begin{figure}[hbtp] 
\unitlength 1mm
\begin{minipage}[t]{5.7cm}
\begin{picture}(57,80)
\put(-4,-2.5){
\epsfig{file=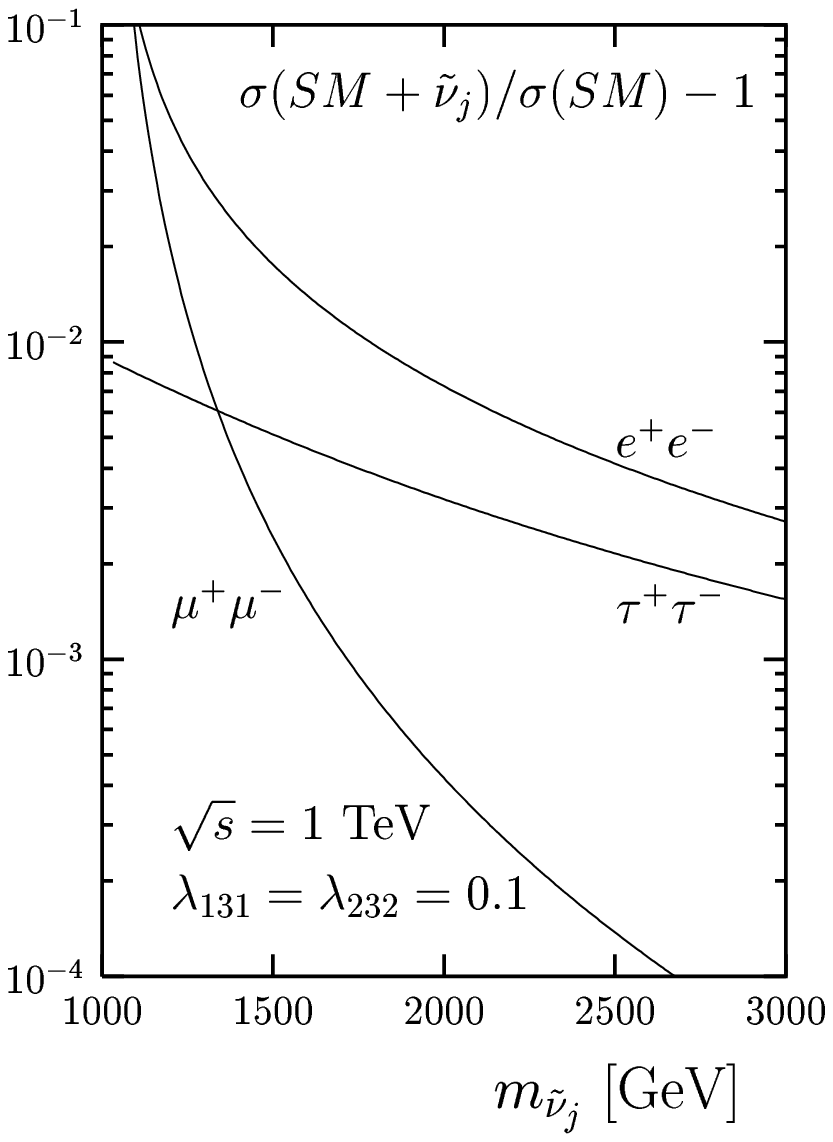,height=8.2cm,width=6.2cm}}
\end{picture}\par
\caption{\it \footnotesize Virtual effects from sneutrino exchange 
  in $e^+e^- \rightarrow l^+l^-$.}
\label{fig:virtual}
\end{minipage}\hfill
\begin{minipage}[t]{6.8cm}
\begin{picture}(68,89)
\put(-1,0){
\epsfig{file=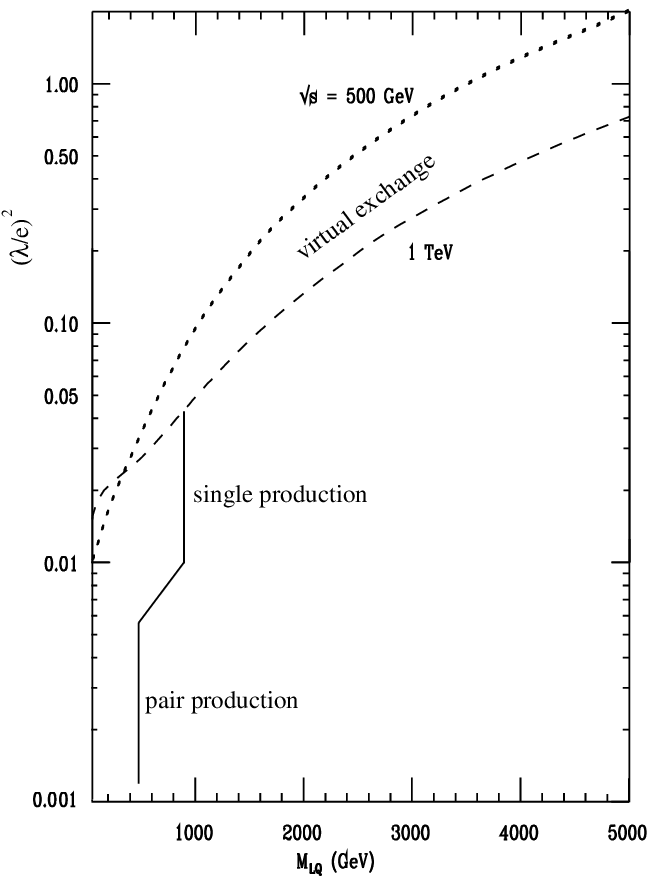,height=9cm,width=6.75cm}}
\end{picture}
\caption{\it \footnotesize LQ search limits at linear $e^+e^-$
  colliders. The boundaries for virtual LQ exchange are taken from 
  Ref.\ \protect\cite{hewett}.}
\label{fig:hewett}
\end{minipage}
\end{figure}

In order to illustrate the size of the deviations from the standard
model predictions at future linear colliders we extrapolate the study of
Ref.\ \cite{KRSZ97} from LEP2 to LC energies.  The result is shown in
Fig.\ \ref{fig:virtual} for $R$-parity violating couplings
$\lambda_{i3k}$ respecting the present experimental bounds
\cite{dreiner}. As can be seen, the effects (which scale roughly with
$(\lambda/m_{LQ})^2$) are of the order of 1\% and smaller, except within
a mass range of $\pm 250 (750)$ GeV of the $s$-channel resonance in
$e^+e^- \rightarrow \mu^+\mu^- (e^+e^-)$.

\section{Summary}

The search limits for leptoquarks in pair production estimated in this
study are independent of the size of the Yukawa couplings $\lambda$, but
they depend somewhat on the LQ quantum numbers. In many cases, the
kinematic limit $m_{LQ}=\sqrt{s}/2$ is reached within a few percent. In
contrast, in single production one can probe the existence of
leptoquarks with masses roughly two times bigger than in pair
production, however, single production requires sizeable Yukawa
couplings. If the latter couplings are large enough, virtual LQ exchange
will lead to observable effects in the total hadronic cross section for
LQ masses even far beyond the total c.m.s energy.  The complementarity
of the above reactions is illustrated in Fig.\ \ref{fig:hewett}.

There is also very useful complementarity of searches in $e^+e^-$,
$pp~(\bar{p}p)$, and $ep$ collisions. Firstly, $e^+e^-$ and
$pp~(\bar{p}p)$ primarily probe LQ gauge couplings, whereas $ep$ probes
Yukawa couplings.  Secondly, while in hadronic collisions leptoquarks
being colour triplets are produced democratically, in $e^+e^-$
collisions one has a hierarchy of production rates reflecting the
electroweak and spin quantum numbers. Finally, a very clean and flexible
environment for the determination of LQ properties is provided by
$e^+e^-$ collisions. Particularly powerful tools are angular
distributions testing spin and Yukawa couplings, and beam polarisation
discriminating between different chiralities of the latter.  The issues
of mass reconstruction, measurement of Yukawa couplings, and
determination of spin, weak isospin, etc. are discussed in
Ref.~\cite{RSS96} where also some illustrative distributions are shown.

Concerning the effects from virtual sneutrino exchange in $e^+e^-
\rightarrow l^+l^-$, one finds that the existing experimental
constraints on $R$-parity violating couplings still allow for large
deviations from the standard model expectations.  In particular, the
occurrence of a $s$-channel resonance in the LC energy range is a
spectacular possibility to look for.


\end{document}